\documentstyle[aps,pra,multicol,epsf]{revtex}
\begin{document}

\title{Quantum Computing with Atomic Josephson Junction Arrays}
\author{Lin Tian and P. Zoller}
\address{Institute for Theoretical Physics, University of Innsbruck, A-6020 Innsbruck, Austria}
\date{\today}
\maketitle

\begin{abstract}
We present a quantum computing scheme with atomic Josephson
junction arrays. The system consists of a small number of atoms
with three internal states and trapped in a far-off resonant
optical lattice. Raman lasers provide the ``Josephson'' tunneling,
and the collision interaction between atoms represent the
``capacitive'' couplings between the modes. The qubit states are
collective states of the atoms with opposite persistent currents.
This system is closely analogous to the superconducting flux
qubit. Single qubit quantum logic gates are performed by
modulating the Raman couplings, while two-qubit gates result from
a tunnel coupling between neighboring wells. Readout is achieved
by tuning the Raman coupling adiabatically between the Josephson
regime to the Rabi regime, followed by a detection of atoms in
internal electronic states. Decoherence mechanisms are studied in
detail promising a high ratio between the decoherence time and the
gate operation time.
\end{abstract}

\begin{multicols}{2}
\section{Introduction}
Josephson effects originate from a tunneling of the particles in
the condensed modes between two superfluids and reflect the phase
difference of the macroscopic wave functions between the
superfluids. Initially discovered in the superconductors
\cite{Tinkham_Superconductivity,Orlando_applied_superconductivity},
Josephson effects have been studied intensively in trapped atoms
both theoretically and experimentally
\cite{Leggett_RMP_BEC,atomic_JJ_exp}. In the atomic case,
Josephson junctions can be constructed between two superfluids
spatially separated by a double well potential and can be
constructed between atomic internal modes coupled by lasers.
Studies include the macroscopic quantum coherence between two
atomic condensates and the observation of the Josephson dynamics.
\cite{bec_JJ_two_well}

One important application of the Josephson junctions discussed in
recent years is in quantum computing.  Various superconducting
Josephson devices have been proposed for implementing a quantum
computer, including the charge qubit, the flux qubit, the phase
qubit.  These qubits have been experimentally tested and have shown
quantum coherent oscillations between macroscopically
distinguishable states\cite{charge_qubit,pc_qubit_1,pc_qubit_2}.

The atomic Josephson junctions can also be explored for quantum
computing. In this paper, we present a candidate for implementing
an atomic ``flux'' qubit with small number of atoms in an optical
trap. We assume that a Bose Einstein condensate with three atomic
states is stored in the lowest vibrational state of an optical
trap\cite{Schulze_thesis_1999}.  The three internal atomic states correspond to three
bosonic modes. Each mode is the analogue of a superconducting
metallic island. Raman lasers generate the Josephson links between
the internal modes, while atomic collisions provide an effective
capacitive couplings between the modes. The  phase differences
between lasers plays the role of the magnetic field in the
superconducting loop. With the competition between the Josephson
energy and the collision energy, the atoms behave collectively and
the stationary states of the qubit have a coherent particle
transfer --the persistent current -- between the internal modes.
With only $15$ atoms\cite{Garcia_Cirac_unknown_parameters_qc}, the atomic counterpart of the
superconducting flux qubit\cite{pc_qubit_1,pc_qubit_2} can be
realized, which bears all the qualitative features of the
superconducting flux qubit.

Compared with the superconducting flux qubit, the parameters of
the atomic ``flux'' qubit can be controlled with large flexibility
and high uniformity. Both the Josephson coupling and the collision
interaction can be adjusted by external electro-magnetic sources.
The Josephson couplings of different junctions can be made to high
accuracy with the fine control of laser. While for
superconductors, not only that the junction  parameters fluctuate
due to the inaccuracy in fabrication, but the parameters are fixed
for one sample. This advantage makes it easier to scale up the
number of qubits in the atomic systems and provides various ways
to implement gate operations. Another merit of the atomic qubit is
that a projective measurement can be performed by adiabatically
switching the Raman couplings. On the contrary, an efficient
readout for the solid-state qubits is a problem many people are
studying. The drawback of the atomic qubit is the slow gate speed
which is limited by the strength of the collision interaction.
Meanwhile, this drawback can be compensated by the long
decoherence time.  In the solid-state systems, various elementary
excitations can damage the coherence of the quantum states in a
time that is only one order longer than the gate time; while we
show that in the atomic qubit, the decoherence time is thousand
times of the gate operation time.

In the following, the major results are summarized. In section II,
we briefly review the superconducting flux qubit and the
experimental achievement for the flux qubit. In section III, we
give a detailed description of our proposal for the atomic
``flux'' qubit  and how it can be realized experimentally. We also
characterize the qubit at different parameter regimes and present
typical energy scales for the qubit. In section IV, we introduce
the phase mode to compare this qubit with the superconducting one
and show that a small number of atoms indeed represents the
macroscopic behavior  of a Josephson junction. This section is
followed by section V where the implementation of quantum logic
gates are studied. In section VI, a quantum nondemolition
measurement scheme is constructed via the adiabatic switching of
the Josephson couplings. The decoherence of the qubit is discussed
in section VII. The  conclusions are given in section VIII.

\section{The superconducting flux qubit}
Josephson junctions have been proved to be a promising building
block for quantum computers. Various proposals of Josephson
circuits at different parameter regimes have been
studied\cite{charge_qubit,pc_qubit_1,pc_qubit_2}. Among these, the
superconducting flux qubit---also named the persistent-current
qubit---has been intensively studied both theoretically and
experimentally.   In the following, we briefly summarize the basic
facts of the flux qubit in superconducting Josephson Junctions, to
allow comparison with the atomic ``flux'' qubit introduced in the
following section.

\subsection{The Circuit of the Qubit}
\begin{figure}[tbh]
\centerline{\epsfxsize=3in\epsfbox{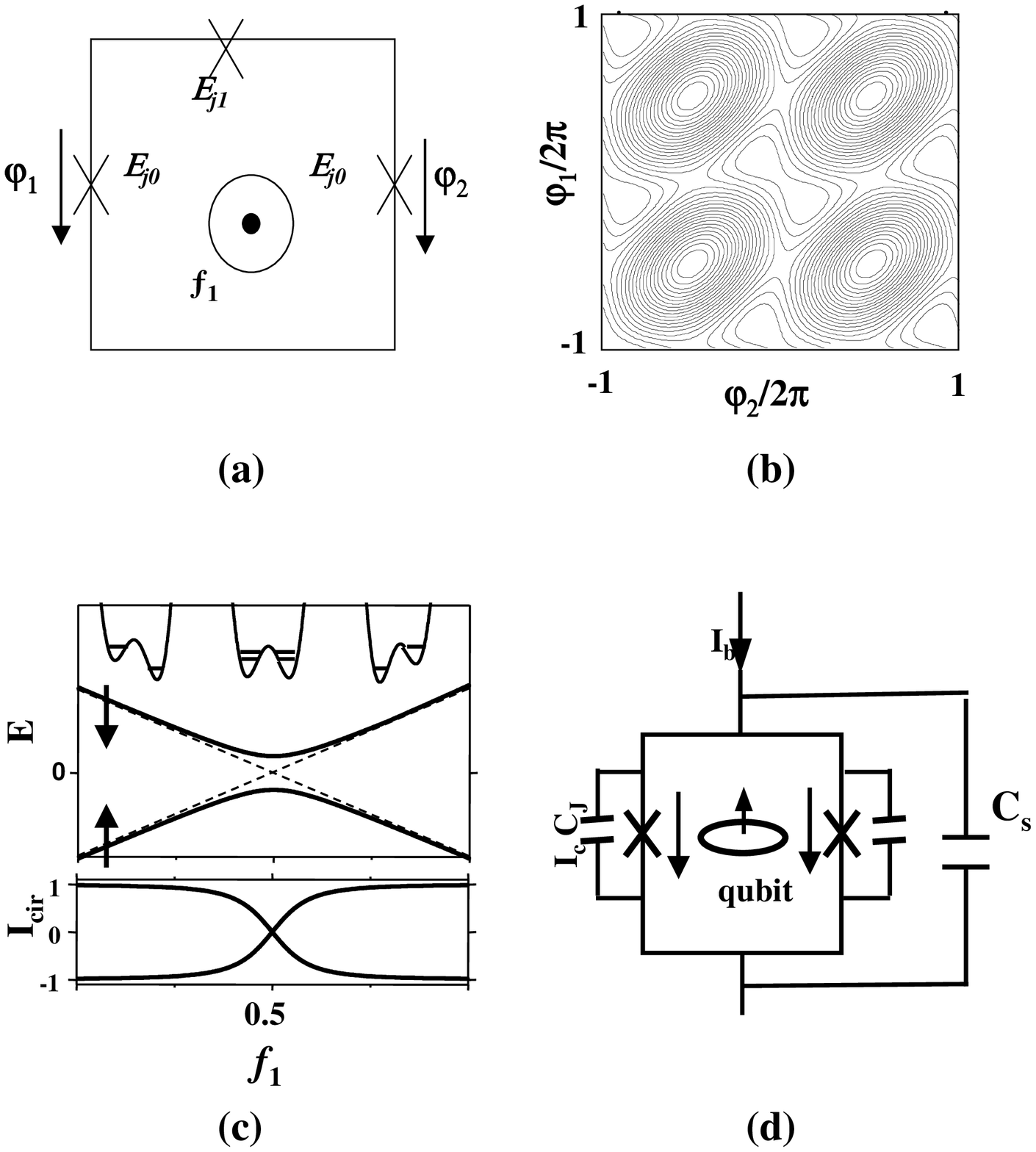}}
\caption{\narrowtext The superconducting flux qubit. (a) the
circuit of the flux qubit. (b) the potential energy for the qubit.
The black centers are local maxima and the white centers are local
minima. (c) the energy and the average current of the qubit versus
the flux. The arrows indicate the qubit states with opposite
currents. The double well potentials at the corresponding flux are
plotted. (d) the measurement of the qubit by a dc SQUID.}
\label{Figure_1}
\end{figure}

The superconducting flux qubit\cite{pc_qubit_1,pc_qubit_2} is a
superconducting loop with three Josephson junctions in series, as
in Fig.~\ref{Figure_1}a. Written in terms of the phase differences
cross the bottom two  junctions, $\varphi_1$ and $\varphi_2$, the
Hamiltonian (Eq. (11) in \cite{pc_qubit_2}) is
\begin{eqnarray}
{\cal H}_t&=&\frac{1}{2}\vec{P}^T{\bf M}^{-1} \vec{P}+
E_J\{ 2+\alpha-\cos\varphi_1-\cos\varphi_2  \nonumber \\
 &&-\alpha\cos(2\pi f_1+\varphi_1-\varphi_2)\}\label{H_pc_qubit}
\end{eqnarray}
where $\vec{P}=(\hat{P}_1, \hat{P}_2)^T$ are the conjugates of the
phase variables and have the physical meaning of the charges on
the islands. The first term is the capacitive energy with ${\bf
M}=(\hbar/2e)^2{\bf C}$ where ${\bf C}$ is the capacitance matrix
of the circuit. The rest of the terms are the Josephson energy
with $E_J=I_c(\frac{\hbar}{2e})^2$ and $I_c$ being the critical
current of the junctions. The third junction at the top of the
circuit has a Josephson energy of $\alpha E_J$ with $\alpha=0.75$.
The magnetic flux in the loop, $f_1$ in unit of the flux quantum
$\Phi_0=\hbar/2e$, is an important control parameter for the
qubit. Both the stationary states and the one-bit logic gates are
controlled via this flux.

The Hamiltonian in Eq.~(\ref{H_pc_qubit}) describes a phase
particle in a two-dimensional periodic potential as is shown in
Fig.~\ref{Figure_1}b. Each unit cell has two energy minima and is
a double well potential. At $f_1=0.495$, the lowest two states of
the qubit localize in one of the two wells respectively and have
opposite circulating currents.  At $f_1=1/2$, the lowest two
states are symmetric and antisymmetric superpositions of the
localized flux states and the energy splitting is due to the
tunneling of the flux states over the potential barrier.
Considering only the lowest states, the qubit can be described by
the Pauli matrices for a $1/2$ spin: $ {\cal
H}_q=\frac{\epsilon_0}{2}\sigma_z+\frac{t_0}{2}\sigma_x$, where
the eigenstates of $\sigma_z$ are the localized flux states and
$\epsilon_0$ varies linearly with $(f_1-1/2$). Typically, the
Josephson energy is $E_J=200\,{\rm GHz}$ and $E_J/E_C=80$.
Numerical calculations of the energy and current are shown in
Fig.~\ref{Figure_1}c. The energy difference of the qubit states at
$f_1=0.495$ is $\omega_q\sim 10\,{\rm GHz}$ with the average
currents of $\pm 0.7I_c$; at $f_1=1/2$, $t_0=10\,{\rm GHz}$.

For a quantum circuit to be a good qubit for fault-tolerant
quantum computing, five requirements have to be
met\cite{DiVincenzo_5_requirements}: 1. to identify a scalable
quantum system; 2. to perform universal quantum logic gates; 3. to
prepare the initial state; 4. to read out the qubit states; 5. to
have a decoherence time longer than $10^4$ quantum operations. The
three-junction loop behaves as an effective two-level system and
can be mapped onto a $1/2$-spin. The qubit can be prepared to the
ground state by cooling it to a temperature of $T\sim 50\,{\rm
mK}\ll\omega_q$.

\subsection{Quantum Logic Gates}
To achieve universal quantum logic operations, two elementary
gates are required: single qubit rotation and two qubit controlled
gate.  For the superconducting flux qubit, the single qubit gate
is by applying microwave oscillations to the superconducting loop.
Typically, the Rabi frequency is $\omega_r=10-100\,{\rm MHz}$ in
proportional to the amplitude of the microwave.  The two qubit
gate is constructed via the  coupling of the circulating currents
of the two qubits: ${\cal H}_{int}=M_{12}|\langle
I_1\rangle\langle I_2\rangle|$ with $I_{1,2}$ being the currents
of the two qubits and $M_{12}$ being the mutual inductance. The
interaction can be of order of $1\,{\rm GHz}$.

\subsection{Qubit State Readout}
The qubit is measured by inductively coupling the qubit to a dc
SQUID magnetometer which is a superconducting loop with two
Josephson junctions as is shown in Fig.~\ref{Figure_1}d. When the
current that flows through the SQUID increases, the SQUID stays in
the superconducting state until a critical current $I_c^{eff}$
where the SQUID makes a transition to a finite voltage state.  The
critical current is varied by the flux generated by the qubit:
$\delta I_c^{eff}=\pm\delta\varphi_q I_c^{sq}\sin\pi f_{ex}$ where
$f_{ex}$ is the external flux in the SQUID and
$\pm\delta\varphi_q$ are flux of the two qubit states
respectively. By measuring the critical current, the qubit states
are read out. Due to fluctuations, the measured critical current
has a distribution that is wider than $\delta I_c^{eff}$ which
results in a nonprojective measurement of the qubit.

\subsection{Decoherence}
Many factors can result in quantum errors against the
superconducting qubits. First, the errors can come from the
imperfect control of the qubit circuits, for example, off-resonant
transitions during gate operations and unwanted dipolar couplings
between qubits. These errors can be prevented by the quantum
control approach. Second, the fluctuations of the environment of
qubit can cause decoherence of the qubit. In the solid-state
qubits, many elementary excitations exist that can damage the
qubit state, including the dipolar interactions between the qubit
and the nuclear spins, the background charge fluctuations, and the
noise coupled to the qubit from the measurement circuits.  The
decoherence time measured in experiments is $100\,{\rm
nsec}$\cite{pc_qubit_exp} which is about $10$ times of the
operation time. This gives a lower bound for the generic
decoherence of the qubit.

\section{The Atomic ``flux'' Qubit}
In this section we present an atomic counterpart of the
superconducting flux qubit.  The qubit is made of a mesoscopic
Bose Einstein condensate of three-level atoms trapped in the
lowest motional states of an optical trap, and interacting with
each other via cold collision. Josephson junctions, which are the
building block of this qubit, are constructed by laser coupling of
the three bosonic modes of the trapped atoms.

\subsection{The Physical System and the Hamiltonian}
\begin{figure}
\centerline{\epsfxsize=3in\epsfbox{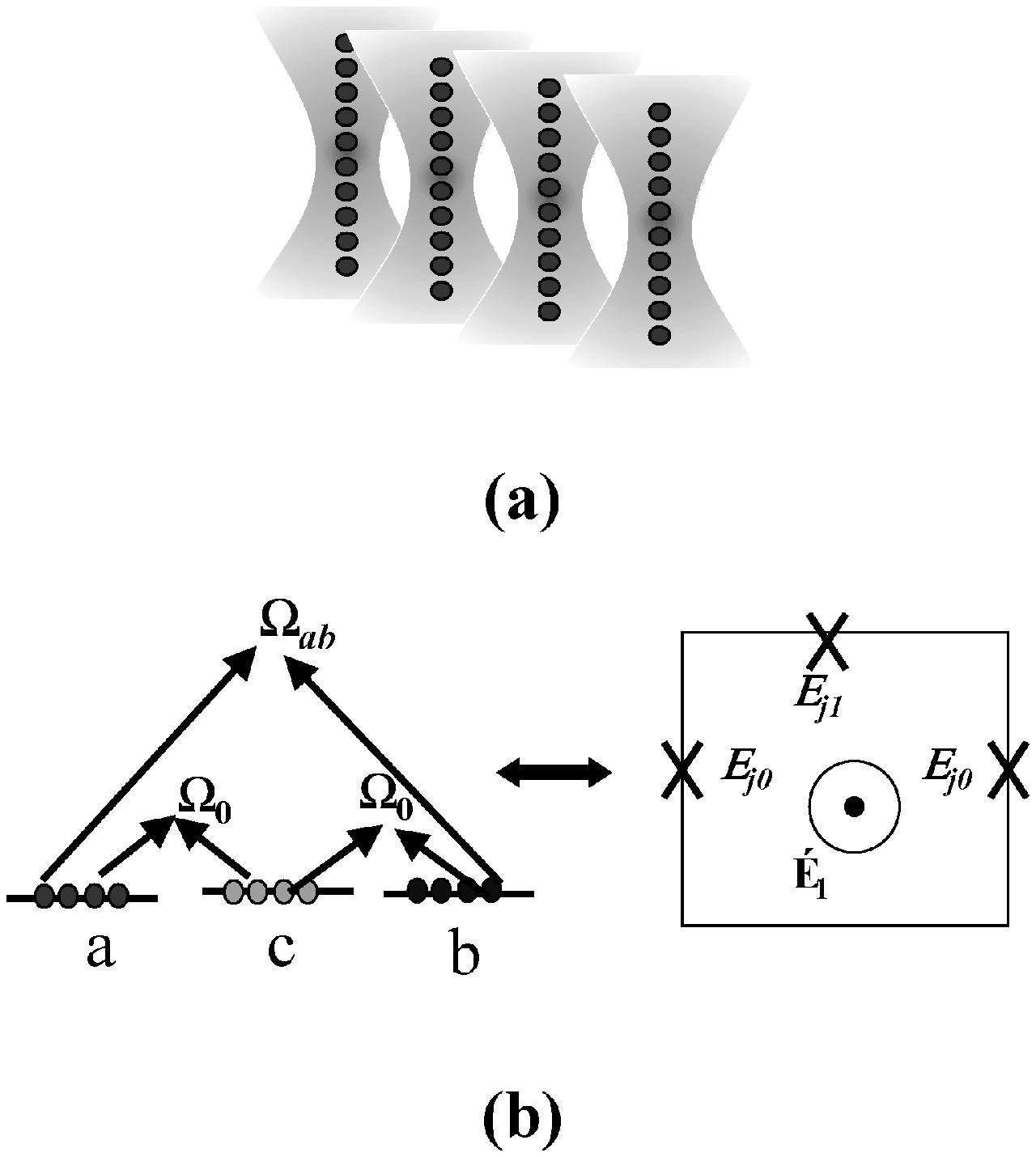}}
\caption{\narrowtext The atomic Josephson junction qubit. (a)
Atoms trapped in the cigar-shaped optical potential by laser
beams. (b) Left: the internal modes coupled by Raman pulses.
Right: the superconducting flux qubit.} \label{Figure_2}
\end{figure}

We consider a small number of three-level atoms trapped in a 1D
optical lattice, as is shown in Fig.~\ref{Figure_2}a, which is
described by the Hamiltonian
\begin{eqnarray}
{\cal H}_0 &=&\displaystyle\sum_\alpha\int
d\vec{x}\,\psi_\alpha^\dagger(\vec{x}) \left
(-\frac{\bigtriangledown^2}{2M}+V(\vec{x})\right )\psi_\alpha(\vec{x})\nonumber \\
&+& \displaystyle\sum_{\alpha,\beta}
U_{\alpha\beta\alpha^\prime\beta^\prime}\int
d\vec{x}d\vec{x}^\prime\,\psi_\alpha^\dagger(\vec{x})
\psi_\beta^\dagger(\vec{x}^\prime)\psi_\beta^\prime(\vec{x}^\prime)
\psi_\alpha^\prime(\vec{x})  \nonumber \\ &+&
\displaystyle\sum_{\alpha\ne\beta} \int d\vec{x}\,\left
(\Omega_{\alpha\beta}(\vec{x})
\psi_\alpha^\dagger(\vec{x})\psi_\beta(\vec{x}) + h.c\right
),\label{H_0_qubit}
\end{eqnarray}
where the three internal states are labeled by $\alpha,\beta$. The
first term is the single particle energy in a harmonic trapping
potential: $V(\vec{x})=\frac{1}{2}m\omega_\parallel^2 x^2+
\frac{1}{2}m\omega_\perp^2 (y^2+z^2)$, where
$\omega_{\perp,\parallel}$ are the trapping frequencies in the
transversal direction and the longitudinal direction respectively.
With a cigar-shaped geometry, we have
$\omega_\perp\gg\omega_\parallel$. We assume that the trapping
frequencies are much larger than all other relevant time scales
(e.g. the qubit energy and the gate speed) so that the atoms stay
in the motional ground states and the qubit can be described by a
three-mode Hamiltonian. The second term in Eq.~(\ref{H_0_qubit})
is the collision interaction.  In practice, this term can be more
complicated due to the scattering between the atoms in different
internal states: for example, in the case of a hyperfine $F=1$,
$M_F=0,\pm 1$ there will be collisional terms where two atoms in
state $M_F=0$ collide to produce an atom in states $M_F=+1$ and
$-1$ states. These terms can be suppressed by shifting the atomic
levels by external fields, so that energy conservation in the
collision is violated, i.e. these terms average away in the
Hamiltonian. Furthermore, detailed  numerical studies show (see
below) that we can simplify the collisional term to a fully
symmetric interaction with
$U_{\alpha\beta\alpha^\prime\beta^\prime}=\delta_{\alpha\beta}\delta_{\alpha\alpha^\prime}
\delta_{\beta\beta^\prime}U_0$, the main physical properties of
the qubit are well preserved. Here, the interaction strength is
\begin{equation}
U_0=\displaystyle\frac{4\pi\hbar^2a_s}{m}\int d^3\vec{x}\,
|\phi(\vec{x})|^4
\end{equation}
where $\phi(\vec{x})$ is the motional ground state of the trapping
potential and $a_s$ is the $s$-wave scattering length with $m$
being the mass of the atoms.  With fixed number of atoms, the
interaction strength increases with the density of the atoms. The
last term in Eq.~(\ref{H_0_qubit}) is the Josephson couplings
between the internal states generated by Raman transitions as is
labeled in Fig.~\ref{Figure_2}b as $\Omega_{\alpha\beta}$. Both
the amplitudes and the phases of these couplings can be accurately
controlled by adjusting the laser parameters.  In this system, we
let $\Omega_{ac}=\Omega_{bc}=\Omega_0$ and $\Omega_{ab}=\Omega_1
e^{i\phi_0}$, where $\Omega_1/\Omega_0$ ranges between $0.5$ and
$1.5$ and is an important factor for the speed of gate operations.
The phase $\phi_0$ is the analogue of the magnetic flux $f_1$ of
the superconducting qubit and is an effective controlling knob for
the  quantum logic gates.

With the above discussion,  the qubit can be well described by a
three-mode Hamiltonian,
\begin{equation}\label{qubit_H}
{\cal H}_0=U_0\displaystyle\sum_{\alpha,\beta}\hat{N}_\alpha^2
+\displaystyle\sum_{\langle\alpha,\beta\rangle}\left (
\Omega_{\alpha\beta}\hat{a}_\alpha^\dagger
\hat{a}_\beta+\Omega_{\alpha\beta}^{*} \hat{a}_\beta^\dagger
\hat{a}_\alpha\right )
\end{equation}
where $\hat{N}_{\alpha}$ is the number operator for the mode
$\alpha$. Here, the collision energy is the slowest energy scale
which limits the speed of quantum logic gates, while the Raman
couplings can be well controlled by lasers. In practice, Feshbach
resonances can be exploited to adjust the scattering length by
several orders of
magnitude\cite{sodium_scattering_length,lithium_scattering_length}
and the gate speed can be improved.

The basis element in this qubit is to construct atomic Josephson
junctions with small number of atoms.  Atomic Josephson junction
has three distinct parameter regimes\cite{Leggett_RMP_BEC}: (1)
the Fock regime with $U_0\gg \Omega_0N_t$; (2) the Josephson
regime with $U_0N_t^2\gg \Omega_0N_t\gg U_0$; and (3) the Rabi
regime with $\Omega_0N_t\gg U_0N_t^2$. In the Fork regime, the
collision energy dominates over the Josephson coupling and the
eigenstates have fixed number of atoms in each internal state. In
the Josephson regime, the qubit behaves as a phase particle in the
Josephson potential energy. In the Rabi regime, the atoms behave
as noninteracting particles described only by the Josephson
couplings. In a superconducting Josephson junction, the Rabi
regime can never be approached with the enormous number of Cooper
pairs. While for the atomic Josephson junctions, all three regimes
are possible. In this paper, we assume the atomic qubit to be in
the Josephson regime. When compared with a large ensemble of atoms
(say $10^5$ atoms) in a superfluid state where the three-mode
approximation becomes inaccurate during fast gate operation, this
system has the advantage that the three-mode model is robust
against the qubit dynamics.

In the Josephson regime, with $N_t\gg1$, Eq.~(\ref{qubit_H}) can
be approximated by a phase model\cite{Leggett_RMP_BEC}.  We
introduce the phase variable $\varphi_{a,b}$ that are the
conjugate operators of the number operators $\hat{N}_{a,b}$
respectively. Due to particle number conservation, $\hat{N}_c$ is
not an independent operator with
$\hat{N}_c=N_t-\hat{N}_a-\hat{N}_b$. By neglecting terms of order
$1/\sqrt{N_t}$, the Hamiltonian is
\begin{eqnarray}
{\cal H}_{phase}&=&-\displaystyle\sum_{\alpha,\beta}
U_{\alpha\beta}\frac{\partial^2}{\partial_{\varphi_\alpha}
\partial_{\varphi_\beta}} +\displaystyle\frac{2}{3}N_t
\Omega_0(\cos{\varphi_a}+\cos{\varphi_b}) \nonumber \\
&+&\displaystyle\frac{2}{3}N_t\Omega_1\cos{(\varphi_a-\varphi_b+\phi_0)},
\label{H_phase_model}
\end{eqnarray}
which shows that with large number of atoms in the qubit, the
Hamiltonian in the phase model maps exactly to
Eq.~(\ref{H_pc_qubit}) of the superconducting flux qubit with
$E_c=3U_0/4$\cite{mapping_qubits}, $E_J=2\Omega_0 N_t/3$. In the
next section, we will discuss the validity of the phase model with
small number of atoms.

We illustrate our model with the following parameters for $15$
sodium atoms in the trap. For $^{23}{\rm Na}$, we choose the trap
size to be $L_\parallel=0.85\,{\rm \mu m}$ and $L_\parallel=10
L_\perp$, which can be achieved with a far red detuned laser. 
The trapping frequencies
are $\omega_0^\parallel =3.7\,{\rm kHz}$ and
$\omega_0^\perp=370\,{\rm kHz}$.  Let $a_s=10\,{\rm nm}$. With a
density of $\rho=3\times 10^{14}\, {\rm cm^{-3}}$, the collision
interaction is $U_0=550\,{\rm Hz}$. The Josephson couplings can be
controlled so that $\Omega_{\alpha\beta}\sqrt{N_\alpha N_\beta}\gg
U_0$ in analogy to the superconducting flux qubit. We let
$2\Omega_0\langle N_\alpha\rangle\approx 70 U_0$ in the following
calculations.  In the notations of the superconducting qubit,
$E_J/E_c\approx 90$.

\subsection{Effective Two Level System}
\begin{figure}
\centerline{\epsfxsize=2in\epsfbox{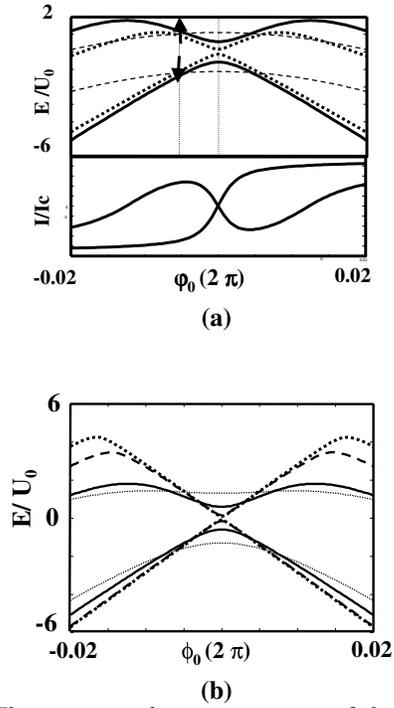}}
\caption{\narrowtext The energy and average current of the qubit
states versus the phase $\phi_0$. (a) $N_t=15$ atoms.   Solid
lines: for symmetric interaction with $U_0=550\,{\rm Hz}$;  dashed
lines: for symmetric interaction with $U_0=0$; dotted lines:  for
interaction given in the text. (b) Energies of the qubit with
various number of atoms for the symmetric interaction. Solid
lines: $N_t=15$; thin dotted lines: $N_t=10$; dashed lines:
$N_t=30$; dotted lines: $N_t=50$. } \label{Figure_3}
\end{figure}

We have numerically studied the Hamiltonian in Eq.~(\ref{qubit_H})
with the above parameters. In Fig.~\ref{Figure_3}a, we plot the
energies and the average currents of the eigenstates of the qubit
versus the phase $\phi_0$ in the range $0.48$ to $0.52$ (in unit
of $2\pi$). The current operator is defined as
$\hat{I}_{ac}=i\Omega_0(\hat{a}_a^\dagger \hat{a}_c
-\hat{a}_c^\dagger\hat{a}_a)$.  We define the lowest two states as
the effective two level system of a qubit. At $\phi_0=0.495$, the
qubit energy is $\omega_q=1.4\,{\rm kHz}$, where the states are
labeled by the arrows. The stationary states have a coherent
transfer of the atoms between the internal states that provides a
persistent particle current for the qubit. The currents of the two
qubit states flow in opposite directions just as in the
superconducting qubit, with $\langle I_{ac}\rangle_{1,2}=\pm 4.3
\Omega_0$. This shows that the atoms behave collectively just as
the electrons in the superconducting wires, which is a result of
the interaction between the atoms. For comparison, the energies of
the qubit when $U_0=0$ are also plotted as the dashed lines in
Fig.~\ref{Figure_3}a.

At $\phi_0=1/2$, the energy splitting $t_0$ is $660\,{\rm Hz}$
which is the counterpart quantum tunneling in the flux qubit and
an important feature of the qubit that is crucial for gate
operations. We studied this splitting with various circuit
parameters. Our result shows a dramatic dependence of $t_0$ on the
ratio between the Josephson couplings $r_0=\Omega_1/\Omega_0$: at
$r_0=0.75$, $t_0=1.1\, {\rm kHz}$; at $r_0=1$, $t_0=0.2\, {\rm
Hz}$; and for $r_0>1$, $t_0$ is almost unchanged as $r_0$
increases.

In this study, we choose the fully symmetric interaction in
Eq.~(\ref{qubit_H}) for the comparison with the superconducting
qubit.  It can be shown that the detailed form of the interaction
doesn't change the main feature of the qubit.  For example, let
the interaction take on the form $V=c_2{\bf\rm \vec{F}_1\cdot
\vec{F}_2}$, where $c_2=(g_2-g_0)/3$ and $g_i=4\pi\hbar^2 a_i/M$
is the scattering strength in the ${\rm F_t=i}$ channel, with
$\vec{F}_i$ being the angular momentum of the atom $i$ and $F_t$
being the total spin\cite{jason_ho_1998}. Numerical results with
$c_2<0$ shows that the calculated energy spectrum, and hence the
properties of the qubit,  has the same butterfly shape as that in
the fully symmetric interaction, as is plotted as the dotted lines
in Fig.~\ref{Figure_3}(a). In this plot, we choose the interaction
to be $-U_0$ with $\Omega_1/\Omega_0=0.85$.

\subsection{Finite ``Size'' Effect}
To see the effect of the small number of the atoms, we calculate
the energies with various numbers of atoms, as is shown in
Fig.~\ref{Figure_3}b. The plot shows that the energies of the
qubit converge as the number of atoms increases. Furthermore, it
shows that when $N_t=15$ the states of the qubit well represents
the key features of a superconducting flux qubit---the features of
a qubit in the phase model. The surprising fact is that with a
small number of atoms, the atomic qubit reflects the properties of
the flux qubit with over $10^{10}$ Cooper pairs: the qubit states
have opposite persistent currents; the phase in the Raman coupling
induces energy difference that is nearly linear with $\phi_0-1/2$;
besides, even the wave functions in the phase space can be
described by the localized phase states.

\begin{figure}[t]
\centerline{\epsfxsize=2.5in\epsfbox{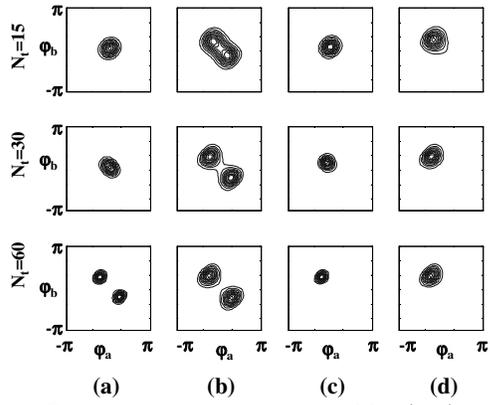}}
\caption{\narrowtext Contour plots of the probability
$|\psi_g(\varphi_a,\varphi_b)|^2$ of the ground state wave
function of the qubit: (a) $U_0=0.01$, $\phi_0=1/2$; (b)
$U_0=0.2$, $\phi_0=1/2$; (c) $U_0=0.01$, $\phi_0=0.495$; (d)
$U_0=0.2$, $\phi_0=0.495$.} \label{Figure_4}
\end{figure}

The wave function in the basis of the phase variables is
$\vert\psi\rangle=\int
d\varphi_ad\varphi_b\vert\varphi_a,\varphi_b\rangle\langle\varphi_a,\varphi_b\vert\psi\rangle$.
In our calculation, we use the number state basis for the states:
$\vert\psi\rangle=\sum_{n_a,n_b} c_{n_a,n_b}\vert n_a,n_b\rangle$,
where $c_{n_a,n_b}$ is the coefficient of the wave function. The
wave function in the phase basis is then:
$\langle\varphi_a,\varphi_b|\psi\rangle=\sum_{n_a,n_b}
c_{n_a,n_b}e^{-i\varphi_a n_a-i\varphi_bn_b}$.  In
Fig.~\ref{Figure_4}, $|\langle\varphi_a,\varphi_b|\psi\rangle|^2$
of the ground state is plotted in the phase basis with
$N_t=15,\,30,\,60$ respectively.

The phase model predicts that at $\phi_0=1/2$ the wave function be
a superposition of two local flux states.  For the small number of
atoms with a weak interaction, Fig.~\ref{Figure_4}a shows that the
qubit state localizes at the center of the phase space in contrast
to the phase model prediction;  while, with $N_t=60$, the state is
a superposition of two local states in agreement with that of the
phase model. Fig.~\ref{Figure_4}c shows the same result for
$\phi_0=0.495$.  With a stronger interaction, Fig.~\ref{Figure_4}b
and Fig.~\ref{Figure_4}d show that the state of $N_t=15$ atoms
agrees with the phase model result. Our study indicates the
behavior of the qubit depends strongly on the factor
$U_0N_t^2/\Omega_0N_t$. When $U_0N_t^2<\Omega_0N_t$, the qubit
enters to the Rabi regime and single atom behavior starts to
dominate over the collective behavior. Our result also shows that
with $N_t=15$, the qubit represents the main features of a
phase-model qubit.

\section{Gate Operations}
Now we are going to discuss how to realize quantum logic gates,
qubit initialization, qubit state readout and the
decoherence of the atomic flux qubit in the following.

\subsection{One-bit Gate}
The superconducting qubit is operated with external magnetic
fields where microwave pulse in resonance with the qubit frequency
is radiated on the superconducting loop. Off-resonant transitions
to other states of the qubit can be neglected since the Rabi
frequency is much smaller than the detuning.

A similar scheme can be applied in the case of the atomic flux
qubit. If we take a Raman laser coupling any two of our bare
atomic states which make up the qubit, and we tune these lasers to
match the energy difference of the qubit states, we can perform
Rabi rotations between the states. In order not to excite any
higher lying states, the Rabi frequency should be less than the
level spacings. In the atomic flux qubit, the qubit frequency and
the detuning are of the order of $1\,{\rm kHz}$ which makes these
gates slow. A first way to improve this, is to use adiabatic
passage, i.e. a sweep of the detuning across the resonance, which
allows a single qubit rotation on the order of the level spacing.
Below we discuss in more detail another scheme based on fast
switching of the phase $\phi_0$ of the Raman coupling
$\Omega_{ab}$.

\begin{figure}
\centerline{\epsfxsize=3in\epsfbox{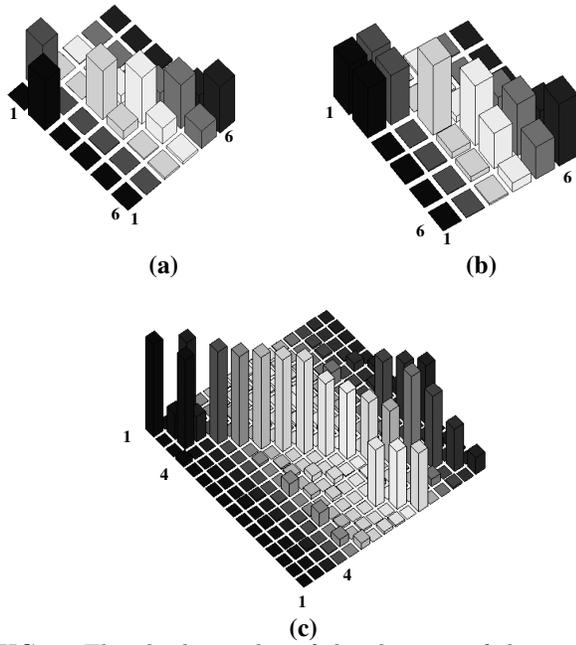}}
\caption{\narrowtext The absolute value of the elements of the
unitary transformations for quantum logic gates, $\vert
U_{ij}\vert$. The transformations are on the lowest six states of
the qubits with the lowest two states the $\vert \uparrow\rangle$
and $\vert\downarrow\rangle$ states of the qubit. (a) single-qubit
NOT gate. The labels indicate the lowest qubit states from $1$ to
$6$. (b) single-qubit Hadamard gate. Labels are same as in (a).
(c) two-qubit gate by a $36$-pulse sequence. The labels $1$ to $4$
are the qubit states: $\vert\uparrow\uparrow, \uparrow\downarrow,
\downarrow\uparrow, \downarrow\downarrow\rangle$. The rest are
higher states.} \label{Figure_5}
\end{figure}

Assume $H_A={\cal H}_0(\phi_0=0.495)$ and $H_B={\cal
H}_0(\phi_0=0.5)$, and $[H_A,\, H_B]\ne 0$. We know from group
theory that by switching the phase alternatively between these two
phase values, any desired unitary transformation can be
constructed within reasonable number of switchings as
$U=e^{-iH_At_{2n}}e^{-iH_Bt_{2n-1}}\cdots
e^{-iH_At_2}e^{-iH_Bt_1}$ by adjusting the durations $t_i$ of the
pulses\cite{Lloyd_PRL_75}. For a single qubit gate, we want the
unitary transformation to be block-diagonal between the two qubit
states and the other states. A numerical optimization of the
$\{t_i\}$ is applied to a $12$-pulse sequence of the $H_A$ and
$H_B$ operators for the lowest six states of the qubit. We
construct a NOT gate and a Hadamard gate $U_h$. The elements of
the unitary operators $|U_{ij}|$ are shown in
Fig.~\ref{Figure_5}a, Fig.~\ref{Figure_5}b.  The off-diagonal elements
$U_{i,1},\,U_{i,2}\ll 0.01$ shows a high fidelity for one-bit
gates.  The total time for the gates is $\tau_1\sim 2\,{\rm
msec}$ for both gates.  The accuracy of the gate can be improved
by increasing the number of pulses in the sequence while keeping
the total gate time short (which means faster switching of the
operators $H_{A,B}$).

\subsection{Two-bit Gate}
Two-bit gates can be constructed by external Josephson tunneling
between neighboring qubits. As we mentioned earlier, external
Josephson tunneling is the tunneling of atoms between spatially
separated condensates.  With the geometry in
Fig.~\ref{Figure_2}a where the qubits are aligned parallel along
the longitudinal direction of the cigar-shaped trap, atoms with
the same internal mode can tunnel from one lattice site to its
neighbor site by switching on a laser pulse for a short time. The
tunneling is enhanced by a factor of $N_t$ of the number of atoms.

We consider the tunneling interaction,

\begin{equation}\label{two_bit_int}
{\cal H}_2=\Omega_t\displaystyle\sum_{\alpha}\left ( a_{1\alpha}^\dagger a_{2\alpha}+
a_{2\alpha}^\dagger a_{1\alpha}\right )
\end{equation}
the index $1$ and $2$ in the operators refer to qubits $1$ and
$2$. The tunneling matrix can be estimated with ${\rm WKB}$
approximation: $\Omega_t\sim
\frac{\omega_\perp}{2\pi}\exp{(-\Delta U/\hbar\omega_\perp)}$,
with $\omega_\perp$ being the plasma frequency of the atoms in the
trapping potential, and $\Delta U$ the trapping barrier for the
qubit. The single particle tunneling $\Omega_t$ is enhanced by the
number of particles and so does the speed of two-bit logic gates.
The tunneling rate can be controlled by adjusting the laser pulse.

The interaction ${\cal H}_2$ can be calculated numerically. The
matrix elements of the operator $(a_\alpha^\dagger)_{ij}$ is
obtained by calculating the overlap between the states $\vert i_{N_t+1}\rangle$ for $N_t+1$
atoms and the states $a_\alpha^\dagger\vert
j_{N_t}\rangle$.  Our calculation shows that this interaction as well as that
of the single-qubit gate induces coupling to the higher states of
the qubits. This problem can be prevented by the same approach as
that of the single-qubit gate---fast pulse sequence to decouple
the lower states from the higher states. We apply a pulse sequence
of $36$-pulses with $H_A={\cal H}_2(\phi_0)$ and $H_B={\cal
H}_0^{(1)}(\phi_0)+{\cal H}_0^{(2)}(\phi_0)$ where ${\cal H}_0^{(1,2)}$
are single qubit Hamiltonian at $\phi_0$. In
Fig.~\ref{Figure_5}c, we show the absolute
values of the matrix elements for the two-bit transformations at
$\phi_0=0.495$ and $\phi_0=0.5$ respectively.  With a total pulse
duration of $2\,{\rm msec}$, the fidelity of the gates for
$N_t=15$ atoms is higher than $99\%$.

\section{Adiabatic Process and Measurement}
The qubit we studied in the previous sections works in the
Josephson regime where $U_0N_t^2\gg\Omega_0N_t\gg U_0$. In this
section, we present a quantum non-demolition measurement scheme
during which the qubit is switched adiabatically between the
Josephson regime and the Rabi regime where $\Omega_0N_t\gg
U_0N_t^2$.  In contrast to the measurement of solid-state
qubit where it takes efforts to build efficient measurement schemes,
our method provides an easy-to-realize way for qubit readout.
The same approach can also be applied to initialize the qubit.

\subsection{Qubit in the Rabi Regime}
In the Rabi regime, when the Josephson energy is much larger than
the collision energy, we neglect the collision term and the qubit
is described by the single atom Hamiltonian,

\begin{equation}\label{H_atom}
{\cal H}_J=\left (\begin{array}{c c c} a^\dagger, & b^\dagger, &
c^\dagger\end{array}\right ) \left (
\begin{array}{c c c} 0  & \Omega_{ab} & \Omega_0 \\
        \Omega_{ab}^{*} & 0 & \Omega_0   \\ 
        \Omega_0  & \Omega_0 &  0
\end{array}
\right )  \left (
\begin{array}{c} a \\  b \\  c \end{array}\right )
\end{equation}
which describes a three-mode atom where the internal modes are coupled by lasers.
The eigenstates can be described by atomic states as

\begin{equation}
{\cal H}_J=\displaystyle\sum_{i=1}^3  \epsilon_i S_i^\dagger S_i
\end{equation}
where $S_i^\dagger$ and $S_i$ are the operators for the atomic
eigenstates and $\epsilon_i$ are the eigenenergies with
$\epsilon_1<\epsilon_2<\epsilon_3$ and $2(\epsilon_2-\epsilon_1)<(\epsilon_3-\epsilon_1)$.
The ground state and lowest
excited states of the qubit with $N_t$ atoms can be described by
the atomic states:

\begin{equation}\begin{array}{ll}
\displaystyle|\psi_1^J\rangle=\frac{(S_1^\dagger)^N_t}{\sqrt{N_t!}}
|0\rangle  & E_1^J=N_t\epsilon_1 \\  \displaystyle|\psi_2^J\rangle=
\frac{S_2^\dagger (S_1^\dagger)^{N_t-1}}{\sqrt{(N_t-1)!}} |0\rangle
& E_2^J=(N_t-1)\epsilon_1+\epsilon_2  \\  \displaystyle|\psi_3^J\rangle= \frac{(S_2^\dagger)^2
(S_1^\dagger)^{N_t-2}}{\sqrt{(N_t-2)!}} |0\rangle & E_3^J=(N_t-2)\epsilon_1+2\epsilon_2 
\end{array}
\label{states_atom}
\end{equation}
where in the ground state $|\psi_1^J\rangle$, all atoms stay in
the lowest atomic state $\vert S_1\rangle$. In the first excited state
$|\psi_2^J\rangle$, one atom is excited to the $\vert S_2\rangle$ state and all
the others stay in the lowest atomic state.  This result is also
confirmed by the numerical calculations.

\begin{figure}[hbt]
\centerline{\epsfxsize=3.5in\epsfbox{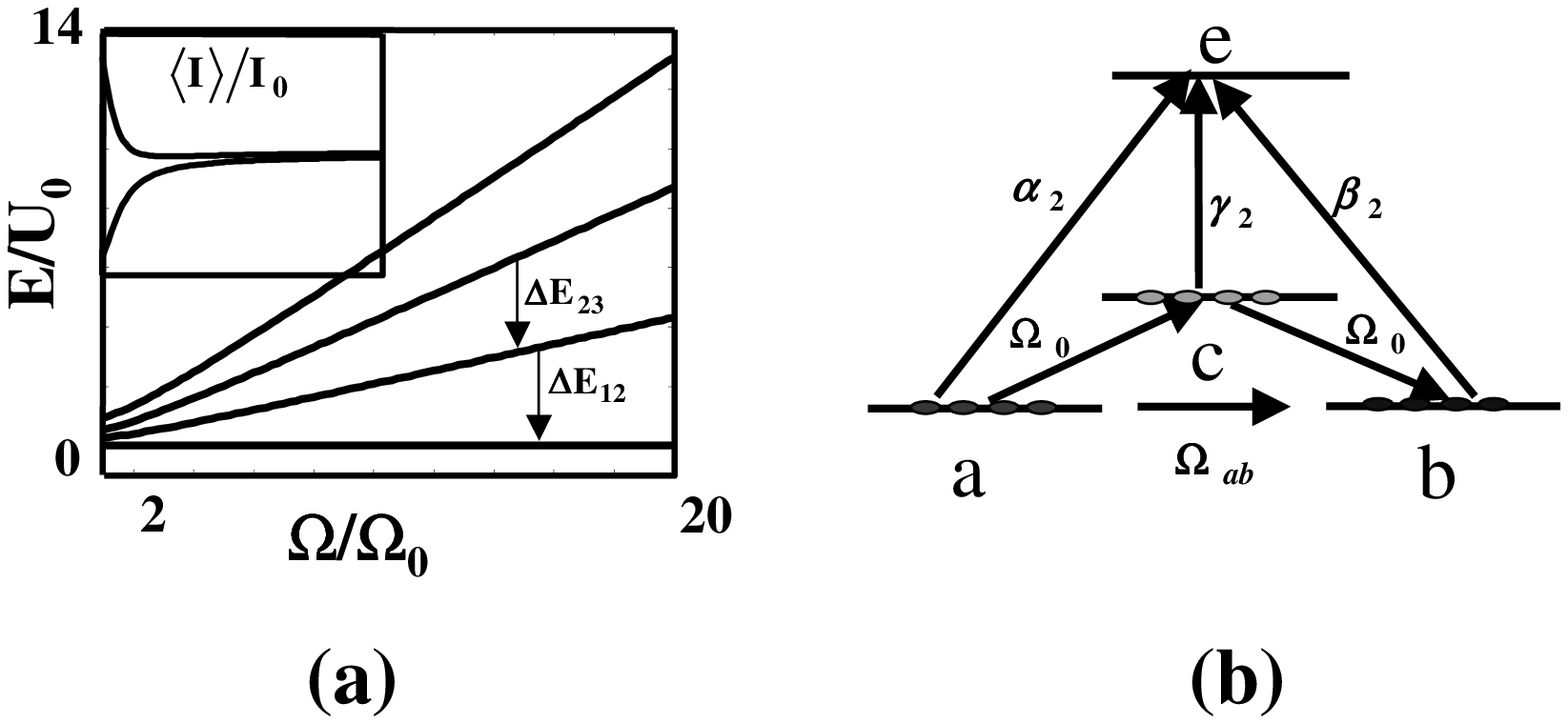}}
\caption{\narrowtext Adiabatic switching of Raman tunnelings. (a)
The energy spectrum of the qubit versus the Raman coupling
$\Omega$. The Raman coupling is plotted in unit of the Raman
coupling $\Omega_0$ for the designed qubit. The energy differences
between states $\Delta E_{12}$ and $\Delta E_{23}$ are indicated
by arrows. The inset shows the average current $\langle
I\rangle/\Omega$ in the same range of $\Omega$. (b) The laser
pulse ${\cal H}_m$ of the QND measurement after the adiabatic
switching. The coupling constants $\alpha_2,\beta_2,\gamma_2$ show
the relative phase between the three components of the pulse.}
\label{Figure_6}
\end{figure}

When the collision term can not be neglected, we numerical solve
the Hamiltonian in Eq.~(\ref{qubit_H}). In Fig.~\ref{Figure_6}a,
the calculated energies for the qubit for a large range of
$\Omega_0$ are plotted.  The inset of this plot shows the
persistent currents of qubit states versus $\Omega_0$. The average
currents $\langle I_{ac}\rangle$  of the two qubit states converge
to each other as $\Omega_0$ increases.

\subsection{Initial State Preparation}
When the Raman coupling $\Omega_0$ is tuned slowly, the qubit
state can be manipulated adiabatically. Here ``slow'' means

\begin{equation}
\displaystyle|\min_{\Omega_0}\{E_2(\Omega_0)-E_1(\Omega_0)\}|^2\gg
\frac{d\Omega_0}{dt}
\end{equation}
where $d\Omega_0/dt$ is how fast $\Omega_0$ is tuned, and
$\min_{\Omega_0}\{E_2(\Omega_0)-E_1(\Omega_0)\}$ is the smallest
energy difference between the qubit states during the tuning
process. As is shown in Fig.~\ref{Figure_6}a, it reaches its
minimum at the left most end when $\Omega_0$ is small. Hence the
switching process takes a time of milliseconds.

This adiabatic process can be exploited for efficiently
initializing the qubit to its ground state.  Starting from the
large $\Omega_0$ limit, we prepare the qubit in its ground state
$|\psi_1^J\rangle$, which is equivalent to preparing all the atoms
in state $\vert S_1\rangle$ and which can be achieved easily. Then,
the Raman coupling is adiabatically decreased to the working regime
so that the ground state $\vert\psi_0^q\rangle$ is reached with
high fidelity.

\subsection{Quantum Nondemolition Measurement}
Second, and most important, the adiabatic switching provides a
scheme for a quantum non-demolition measurement of the qubit.
Starting from the working parameters of the qubit where $2\Omega_0
N_a/U_0=70$ and assuming an initial state $\alpha|\psi_0^q\rangle
+\beta|\psi_1^q\rangle$, $\Omega_0$ is slowly increased to the
Rabi regime. When $\Omega_0 N_a\gg U_0$, the qubit state evolves
to $\alpha|\psi_1^J\rangle +\beta|\psi_2^J\rangle$, a
superposition of the states in Eq.~(\ref{states_atom}). As the
increase of $\Omega_0$ is adiabatic, no transition to the excited
state is induced. Then,  a dark-state measurement scheme is
performed on the qubit. Namely, a laser pulse is applied that
excites the atomic state $|S_2\rangle$ to an excited state
$|e\rangle$ and does nothing to the atoms in the states
$|S_1\rangle$ and $|S_3\rangle$. The state $|e\rangle$ emits a
photon via spontaneous emission which is then detected. As can be
seen from Eq.~(\ref{states_atom}), when the laser is applied to the
ground state $|\psi_1^J\rangle$, no transition happens and no
photon is emitted; when the laser is applied to the second state
$|\psi_2^J\rangle$, one atom is excited to the state $|e\rangle$
and one photon is emitted. Hence, this approach achieves a
projective measurement of the qubit.

The laser pulse applied after the adiabatic switching is

\begin{equation}
{\cal H}_m=e^\dagger S_2 + S_2^\dagger e
\end{equation}
where $e^\dagger$ and $e$ are the operators for the excited state
$|e\rangle$. It is easy to prove that single atom states $S_1$ and
$S_3$ are dark states of this operator that can not be excited by
this pulse (as they are orthogonal states of the Hamiltonian in
Eq.~(\ref{H_atom}). We have: ${\cal H}_m |\psi_1^J\rangle=0$ and
${\cal H}_m |\psi_1^J\rangle=|(N-1)_{S_1},1_e\rangle$. By
performing a single photon measurement with the quantum jump
approach, the probability of the qubit in $|\psi_1^J\rangle$, and
hence in  $|\psi_1^q\rangle$ originally, can be detected. For the
qubit in its ground state, the measurement won't do anything and
is a quantum nondemolition measurement.

\section{Decoherence}
A major obstacle in the pursue of quantum computation with
solid-state qubits is the strong coupling to noise and the
resulting low quality factor. In experiment, the  measured
decoherence time for the superconducting qubits is $T_2=100\,{\rm
nsec}$, while the gate time is $\tau_{gate}=10\,{\rm
nsec}$\cite{pc_qubit_2}.

In the atomic qubit presented in this paper, the quality factor
due to decoherence is is high compared with that of  the
superconducting qubit. The qubit is designed to be insensitive to
the major factors that can result in decoherence. For example, all
the energies involved in qubit operation are much lower than the
trapping frequency in the longitudinal direction of the trap
$\omega_\parallel=3.7\,{\rm kHz}$, which keeps the atoms in the
motional ground state during gate operations. Other factors such
as the inaccuracy in the Raman couplings, the particle loss from
the trap and the spontaneous emissions can be well neglected
within a time of seconds.

The fluctuation of the number of atoms could induce severe qubit
decoherence when the number of atoms is large. For example, the
decoherence rate due to single particle loss grows linearly with
$N_t$ and the decoherence rate due to three body collision increases with $N_t^3$. Our
study shows that for single particle loss process with coupling
constant $\gamma_0$, the decoherence rate at $N_t=15$ is $4\gamma_0$,
and the decoherence due to three body collision can be neglected.

\subsection{Effect of Single Atom Loss}
Consider for example a single atom loss characterized by a loss
rate $\gamma$. The time evolution of the density matrix is
described by the following master equation
\begin{equation}\begin{array}{c}
\displaystyle\frac{\partial\rho^t}{\partial t}=-i[{\cal H}_I(t),\,
\rho^t], \\  \displaystyle\frac{\partial\rho^t}{\partial
t}=-\gamma_0\displaystyle\sum_\alpha \left (
\hat{a}_\alpha^\dagger\hat{a}_\alpha\rho^t+\rho^t\hat{a}_\alpha^\dagger\hat{a}_\alpha
-2\hat{a}_\alpha\rho^t\hat{a}_\alpha^\dagger\right ),\end{array} \label{density_matrix}
\end{equation}
where $\rho^t$ is the density matrix of the qubit in the
interaction picture and the atomic losses in different modes are
summed up.

The density matrix can be decomposed into the Hilbert spaces of
different number of atoms: $\rho^t=\sum_n\rho_{ij}^{(n)}\vert
i_n\rangle\langle j_n\vert$, where $\rho_{ij}^{(n)}=\langle
i_n\vert \rho^t\vert j_n\rangle$ is the element of the density
matrix with $n$ atoms and $\vert{i_n,j_n}\rangle$ are qubit states
of $n$ atoms.  Substituting this expression into
Eq.~(\ref{density_matrix}) and assuming an initial density matrix
$\rho^0$ with $N_t$ atoms, we have
\begin{equation}\begin{array}{l}
\rho_{ij}^{eff}(\delta t)=\rho_{ij}^{(N_t)}+\rho_{ij}^{(N_t-1)} \\
[2mm]
\rho^{(N_t)}=\rho^0 -\delta
t\gamma_0\displaystyle\sum_\alpha A_\alpha^\dagger A_\alpha\rho^0
+\rho^0A_\alpha^\dagger A_\alpha \\
\rho^{(N_t-1)}=+2\delta t\gamma_0\displaystyle\sum_\alpha
A_\alpha\rho^0A_\alpha^\dagger
\end{array}
\end{equation}
where the matrix $(A_\alpha^\dagger)_{ij}= \langle i_{N_t}\vert
\hat{a}_\alpha^\dagger\vert j_{N_t-1}\rangle$. Starting from $N_t$
atoms in the trap, when one atom leaks out, the qubit state is a
superposition of the eigenstates of $(N_t-1)$ atoms.  The
decoherence rate is slowed down by the fact that the remaining
system of $(N_t-1)$ atoms largely overlaps with the original qubit
states in the $(N_t-1)$-atom basis.  The decoherence rate is
expressed as
\begin{equation}\label{gamma_eff}
\gamma_{\rm eff}=\gamma_0\displaystyle\max_{ \vert\Psi\rangle}
\{\sum\langle\Psi\vert A_\alpha^\dagger
A_\alpha\vert\Psi\rangle-|\langle\Psi\vert
A_\alpha^\dagger\vert\Psi\rangle|^2\}.
\end{equation}

\begin{figure}[hbt]
\centerline{\epsfxsize=2in\epsfbox{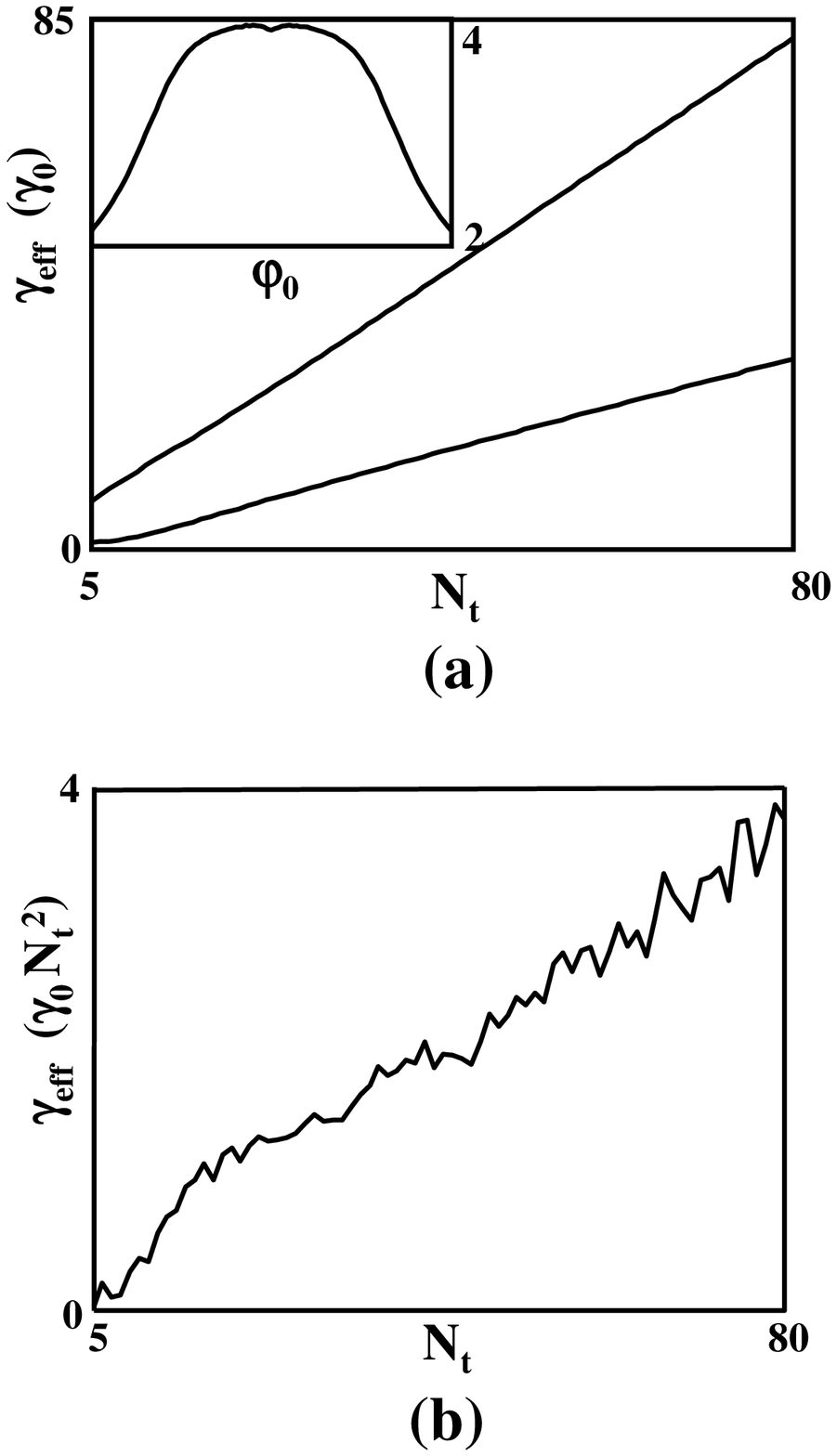}}
\caption{\narrowtext Decoherence rate by Eq.~(\ref{gamma_eff}). (a)
Single atom loss rate versus number of
atoms. Inset: Single atom loss rate versus the phase $\phi_0$ of
Raman coupling $\Omega_{ab}$. (b) Three-body loss rate in unit of 
$\gamma_0 N_t^2$ versus the number of atoms.} \label{Figure_7}
\end{figure}

Numerical results show that $\gamma_{\rm eff}$ grows linearly with
number of atoms in the trap as is plotted in Fig.~\ref{Figure_7}a.
The decoherence is slowed down by a factor of two by the second
term in the above equation. At $N_t=15$, $\gamma^{eff}=4\gamma_0$.
In the inset of Fig.~\ref{Figure_7}a, the dependence of
decoherence on the phase of the Raman coupling $\phi_0$ is plotted
which is flat in the range of interest.

\subsection{Three-body Collision Loss}
One of the main decoherence against this qubit is the three-body
collision loss. The three body process: $A+A+A\rightarrow A_2+A$
describes that  when three atoms collide, two atoms form a bounded
molecular state with a binding energy of order of $\hbar^2/ma_s^2$
which is several orders larger than the trapping frequency, where
$a_s$ is the $s$-wave scattering length. As a result, the molecule
and atom gain very large kinetic energy after the collision and
escape from the trap. This process damages the coherence of the
qubit states. The three-body loss is characterized by
$\gamma_0^{(3)}=K_3(3\pi^3)^{-3/2}\rho^2/72N_t^2$ where $K_3$ is
the three body collision rate in \cite{loss_rate_Esry_1999} and
$\int\,d^3\vec{x}|\psi(\vec{x})|^6$ gives the dependence on
density. We apply the same approach as that in the single atom
loss to calculate the effective decoherence rate and the results
are plotted in Fig.~\ref{Figure_7}b.  It is shown that
$\gamma_{\rm eff}/\gamma_0^{(3)}N_t^2$ grows linearly with $N_t$
and at $N_t=15$, with $\rho=3\times 10^{14}\,{\rm cm^{-3}}$ and
$K_3=10^{-28}\,{\rm cm^6/sec}$, we have $\gamma_{\rm eff}\approx
10^{-4}$ which gives  a very long decoherence time.

\section{Conclusions}
We have presented a scheme for implementing an
atomic  ``flux'' qubit with atomic Josephson junctions, which are
generated by Raman lasers that introduce coupling between internal
modes of atoms. By trapping three internal modes and coupling them
with the Raman pulse, a three junction loop is constructed. The
collision interaction between the atoms provides the analog of the
capacitance energy. With small number of atoms, the qubit presents
the main features of the mesoscopic circuit---superconducting flux
qubit:  the butterfly shaped energy spectrum, the persistent
currents and the local wave function in phase basis.  We have
outlined methods for the implementation of quantum logic gates
with fast switching of Raman pulses, the state initialization, and
we have presented a qubit measurement scheme by adiabatic
switching of the Josephson coupling and observation of quantum
jumps. Furthermore, we have given detailed analysis of possible
imperfection and decoherence of the qubit.

The solid-state qubits suffer severely from noise, which may
become the biggest obstacle in implementing those qubits. However,
the solid-state proposals are easy to scale up and control with
existing technology. The qubit proposed in this paper inherits
many of the merits of the superconducting qubits.  For one thing,
almost all the parameters of the qubit can be very well controlled
by external sources which increases the flexibility of qubit. The
system is in principle scalable by storing the atomic flux qubit
in wells of the 1D optical lattice. Compared with superconducting
qubit, the atomic Josephson junction qubit has the advantage of
not subjecting to severe environmental disturbance and having a
long decoherence time. Hence, an array of the atomic qubits can be
arranged in a space to simulate a ``clean'' array of
superconducting qubits and perform certain quantum gate
operations. Clearly, one of the  main differences to the
superconducting case is the significantly slower time scale of
operations.

In summary, our study shows that the atomic systems can be
designed to be a clean realization of the Josephson junction
circuits and keep the merits of exploring macroscopic/mesoscopic
degrees of freedom and a long decoherence time. In this system,
the Josephson couplings can be controlled with large flexibility
by adjusting the power and phases of the laser beams. The
collision interaction can also be adjusted in a large range by
magnetic-optical means such as tuning around the Feshbach
resonances \cite{Feshbach_resonance}. Moreover, the trap geometry
and the interaction between neighboring qubits can be chosen to
suit different experiments.

\section*{Acknowledgments}
We would like to thank J.I. Cirac for helpful discussions.
Work at the University of Innsbruck is supported by the Austrian
Science Foundation, European Networks and the Institute for
Quantum Information.

\newpage

\newpage

\newpage

\newpage

\newpage

\newpage

\newpage

\end{multicols}

\begin{thebibliography}{99}
\bibitem{mapping_qubits}
Let us introduce $\theta_\pm=(\theta_a\pm\theta_b)/2$. Then the
conjugates are $\hat{P}_\pm= \hat{P}_a\pm\hat{P}_b$.  With
$U_{\alpha\beta}=\delta_{\alpha\beta}U_0$ and $U_{\alpha\beta}$,
the collision interaction is $\frac{3}{2}U_0
\hat{P}_p^2+\frac{1}{2}U_0\hat{P}_m^2$, which shows $E_c=3U_0/4$
when compared with Eq. (12) in \cite{pc_qubit_2}.
\end{thebibliography}
\end{document}